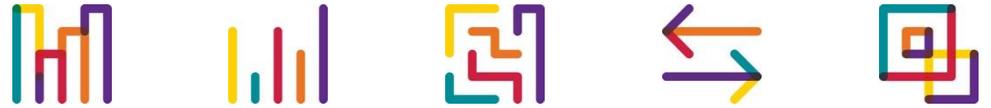

**WHITE PAPER**

# Initial Impacts of COVID-19 on Transportation Systems: A Case Study of the U.S. Epicenter, the New York Metropolitan Area

Authors: Jingqin Gao, Suzana Duran Bernardes, Zilin Bian
NYU PI: Kaan Ozbay, Managing Director: Shri Iyer

Contact: c2smart@nyu.edu
c2smart.engineering.nyu.edu

## Overview

The novel Coronavirus COVID-19 spreading rapidly throughout the world was recognized by the World Health Organization (WHO) as a pandemic on March 11, 2020[1]. This white paper looks at the initial impacts COVID-19 has had on transportation systems in the metropolitan area of New York, which has become the U.S. epicenter of the coronavirus.

As of March 29, confirmed coronavirus cases in the United States surpassed 140,000, and of those, 2,467 people have died[2]. New York has the highest number of confirmed cases (59,513 confirmed cases as of March 29[3]) in the United States, followed by its neighboring state New Jersey with 16,636 confirmed cases[4].

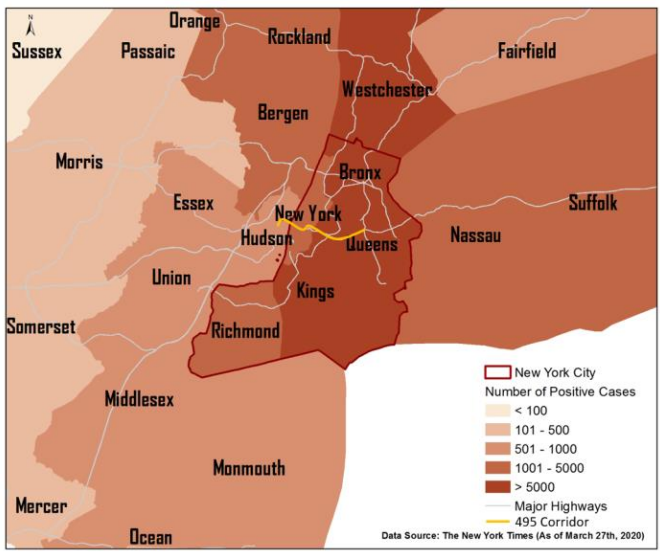

**Figure 1** Positive Cases in the New York Metropolitan Area

With social distancing policies in place, the COVID-19 outbreak has dramatically changed travel behavior. The C2SMART research team at NYU has been investigating temporal changes in travel trends and reviewed the initial impact of COVID-19 on the metropolitan area of New York's transportation system before and after the issuance of stay-at-home orders. Tolled vehicle trips, transit ridership, travel time, weigh-in-motion data, crash rate, and parking citations were analyzed. *Note that all data is preliminary.*

## Timeline

New York State's first case of the coronavirus was reported on March 1st. Since then, several regional and national policies have taken shape in the area impacting the population's transportation. Table 1 shows the timeline of government responses in the metropolitan area of New York.

**Table 1** Government Responses Timeline

| Date | Event |
|---|---|
| Mar 7 | NY state disaster emergency declaration |
| Mar 9 | NJ state disaster emergency declaration |
| Mar 10 | CT state disaster emergency declaration |
| Mar 11 | WHO declared the outbreak a pandemic |
| Mar 13 | National state of emergency declaration |
| Mar 16 | NY/CT: 50+ people gathering banned, Closure of theaters, gyms, casinos, K-12 Schools<br>NJ: 50+ people gathering banned, Closure of theaters, gyms, casinos, Curfew after 8PM |
| Mar 18 | NJ: K-12 Schools closure |
| Mar 20 | NJ: Shut down all non-essential business |
| Mar 21 | NJ: Stay at home order for all residents and canceled gatherings of any number |
| Mar 22 | NY: "New York State on PAUSE"/ closure for all non-essential businesses statewide |
| Mar 28 | CDC issues domestic travel advisory for NY, NJ and CT |

---

[1] "WHO Director-General's opening remarks at the media briefing on COVID-19—11 Mar 2020". World Health Organization. 11 Mar. 2020.
[2] Johns Hopkins University, coronavirus.jhu.edu/map.html

[3] New York State, County by County Breakdown of Positive Cases, coronavirus.health.ny.gov/county-county-breakdown-positive-cases
[4] New Jersey Department of Health, covid19.nj.gov/#live-updates

## The Stay-at-home Effect on Volumes & Ridership

### MTA Facilities

The Stay-at-home orders caused an immediate direct impact on usage of Metropolitan Transportation Authority (MTA) services, the largest public transit authority in the US, which typically carries over 11 million transit riders on an average weekday. Its bridges and tunnels serve more than 800,000 vehicles each weekday and carry more traffic than any other bridge and tunnel authority in the nation[5]. As the stay-at-home order took effect, ridership data show steep declines in both transit ridership and vehicular traffic. The decline rates reached up to 94% in peak transit ridership as of March 23$^{rd}$, and up to 72% in vehicle traffic through MTA bridges and tunnels as of March 29$^{th}$, compared to the same date/week in 2019 (Table 2).

**Table 2** Decline in Transit Ridership and Vehicle Traffic

**Transit Ridership**
*Based on statistics from MTA Annual Disclosure Statement Supplement, March 25th, 2020*

% Change in ridership compared to the same date in 2019

|  | 3/12 | 3/16-3/17 | 3/20-3/23 |
|---|---|---|---|
| NYC Subways | -18.6 | -59.9 | -86.9 |
| NYCT Bus | -14.4 | -48.3 | -60.3 |
| MTA Bus | -18.9 | -51.3 | -62.4 |
| Metro North Railroad (a.m. peak) | -48 | -90 | -94 |
| Long Island Rail Road (a.m. peak) | -31 | -67 | -71 |

**Vehicle Traffic via MTA Bridges and Tunnels** (Data source: MTA[6])

% Change in vehicle traffic compared to the same week in 2019

| Toll Plaza | 3/1-3/8 | 3/8-3/15 | 3/15-3/22 | 3/22-3/29 |
|---|---|---|---|---|
| Robert F. Kennedy Bridge Queens/ Bronx Plaza | -0.1 | -15.5 | -41.7 | -55.2 |
| Robert F. Kennedy Bridge Manhattan Plaza | -2.7 | -17.9 | -48.5 | -63.3 |
| Bronx-Whitestone Bridge | 4.1 | -12.6 | -36.8 | -50.2 |
| Henry Hudson Bridge | 0.3 | -24.9 | -56.7 | -72.3 |
| Marine Parkway-Gil Hodges Memorial Bridge | 8.1 | -9.0 | -33.9 | -49.5 |
| Cross Bay Veterans Memorial Bridge | 4.2 | -6.9 | -29.5 | -38.8 |
| Queens Midtown Tunnel | 0.1 | -20.8 | -52.7 | -68.0 |
| Hugh L. Carey Tunnel | 0.8 | -16.7 | -47.6 | -66.3 |
| Throgs Neck Bridge | -3.6 | -15.4 | -37.4 | -48.2 |
| Verrazano-Narrows Bridge | 2.2 | -10.2 | -32.1 | -43.4 |

The number of positive cases in New York City (NYC), along with the daily changes in subway turnstile entries and tolled traffic on MTA bridges and tunnels, are illustrated in Figure 2.

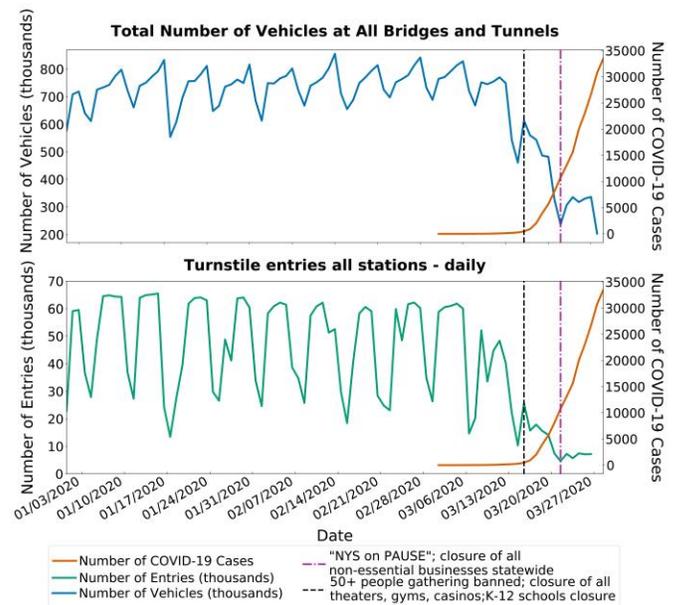

**Figure 2** Daily changes in the number of positive cases in NYC, subway turnstile entries and traffic on MTA bridges and tunnels

The above figure shows the accelerated declines through the first month of the COVID-19 pandemic, particularly after March 16$^{th}$, when new guidelines urging people to avoid social gatherings of more than 50 people were issued and schools, bars and restaurants were closed. Most of the subway stations saw a dramatic decrease in ridership as the week of March 8$^{th}$. For example, Time Square station had a ridership decline of 39.6%, 81.4%, and 92.2% as of the week of March 8$^{th}$, 15$^{th}$ and 22$^{nd}$, respectively. The impact of closing non-essential businesses, March 22$^{nd}$, is also noticeable in the use of both modes.

### Travel Time Trends

The reduction of trips has resulted in a dramatic reduction in average travel times as well. Travel times across the 495 connecting the NJ highway network to the Lincoln Tunnel, across Manhattan and across 34$^{th}$ Street to the Queens-Midtown Tunnel into Long Island, were analyzed in Figure 3.

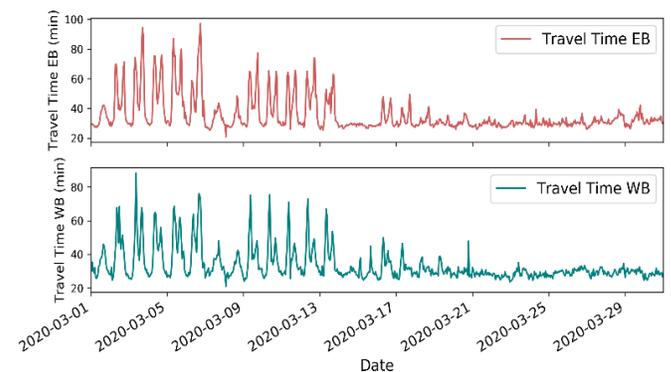

**Figure 3** Travel times on the 495 corridor during COVID-19

---

[5] The MTA Network: Public Transportation for the New York Region, http://web.mta.info/mta/network.htm

[6] NYS Open Data Portal, Hourly Traffic on MTA Bridges and Tunnels.

Overall, the stay-at-home policies have had a significant impact on average travel times on the 495 corridor. In the eastbound direction, travel times averaged 30.2 minutes for the week of March 23rd (the first week of the stay-at-home policies) compared to 48.7 minutes during the 3rd week of February, a 38.0% drop. The decreases during peak hours were even more extreme: travel times were down by an average of 54.3% during AM peak and 61.2% during PM peak on average, compared with February. Figure 4 shows how instead of typical spikes of morning and evening congestion, a flatter travel time pattern is observed, indicating that volumes remain below system capacity throughout the day. Variability in travel time has dropped as well: travel times for Mondays/Tuesdays in February showed a standard deviation of 14.7/19.0 minutes respectively, compared to 6.5, 2.8, and 1.7 minutes on March 16th (50+people gathering banned), 20th (NJ shutdown of all non-essential business), and 23rd (after "NYS on PAUSE").

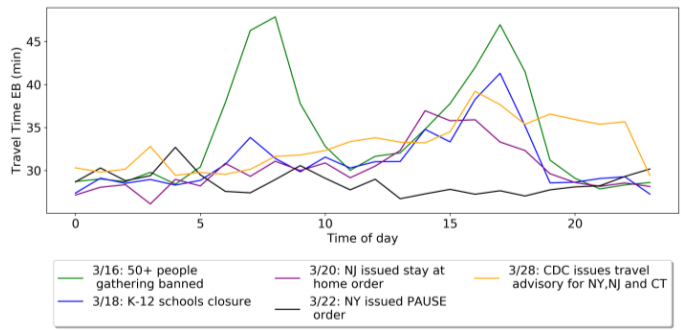

**Figure 4** Travel time changes corresponding to different government orders

### Weigh-in-Motion (WIM) Data from the BQE

C2SMART is tracking truck volumes and weights from weigh-in-motion (WIM) systems installed on its *Urban Roadway Testbed* on the Brooklyn-Queens Expressway (BQE) [7] to observe effects on trucks moving through New York City. Some initial data has been collected on traffic volume, travel speed, and gross vehicle weight (GVW). *Note that this data is preliminary to illustrate trends and is still being validated.*

- **Volume:** After 3/13, total traffic dropped by 30% (Queensbound [QB]), and 29% (Staten Island-bound [SIB]) for the rest of March. However, truck traffic appears to have dropped less than all traffic (15% QB; 19% SIB).
- **Speeds:** The reduction of trips increased average vehicle speeds by 11% (QB) and 24% (SIB). WIM data shows that the typical speed range on this section of the BQE increased by approximately 5mph for both QB and SIB.
- **Gross vehicle weight (GVW):** WIM data does not show a large change to GVWs; however, the number of heavy trucks (>80 kips) as well as the maximum GVW appear to have gone down: 20% reduction for QB and 14% reduction for SIB. This shows that while there are fewer trips, the loads of these trips are also likely lower.

---

[7] http://c2smart.engineering.nyu.edu/2020/04/01/c2smart-roadway-urban-testbed/
[8] Data Source: NYCDOT/Citi Bike

### Mode Shift to Cycling

A temporary mode shift to cycling was observed in the first ten days in March prior to the issuance of the stay-at-home order. Based on NYCDOT data, cycling traffic had a 55% uptick compare to the same period last year over the Brooklyn, Manhattan, Williamsburg and Queensboro bridges[8]. Ridership soon went down due to the stay-at-home order, however Citibike reported that its stations directly adjacent to hospitals exhibited a significant ridership increase[9]. As a result, Citibike launched its Critical Workforce Membership Program.

### Crash Reductions due to Reduced Volumes

According to NYPD's Motor Vehicle Collisions crash reports, while the overall crash rate in NYC was already decreasing in the first three months of 2020 compared to 2019, the decrease after March 16th (with stay-at-home policies in place) was even more significant (Figure 5). Police data shows that motor vehicle collisions were down by an average of 50% in March (8,629 events), compared to the same period in 2019 (17,308 events). Between March 25th and 30th, the number of reported crashes was down by 70-77%, likely due to the reduction in vehicle trips. A 51% reduction in pedestrian injury/fatality and 31% in cyclist injury/fatality in crashes was also observed.

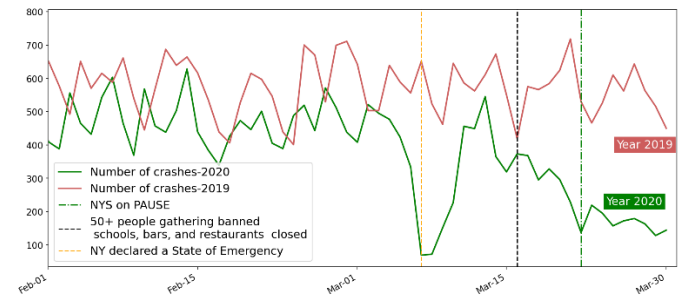

**Figure 5** Number of Crashes during COVID-19 Outbreak

### Parking Citations

The number of parking citations issued in NYC for commercial vehicles[10] were down by 26.1% but remained nearly the same (1.1% increase) for passenger vehicles during the third week of March (3/16 to 3/22) compared to the 3rd week of March in 2019 (Figure 6). Although parking meters remained in effect during the outbreak, right after the issuance of "NYS on PAUSE" on March 22nd, citations were down 74.0% and 64.3% percent for commercial and passenger vehicles, respectively, on March 23-24, compared to the same Monday/Tuesday in the fourth week of March in 2019. Fewer vehicular trips may be contributing to the decrease of citations as more curbside parking spaces became available, but lifting of alternate side parking regulations or other changes in enforcement during the COVID-19 crisis may also be impacting the totals.

---

[9] Citi Bike Critical Workforce Membership Program, www.citibikenyc.com/critical-workforce-membership-press-release
[10] Data Source: NYC Open Data Portal

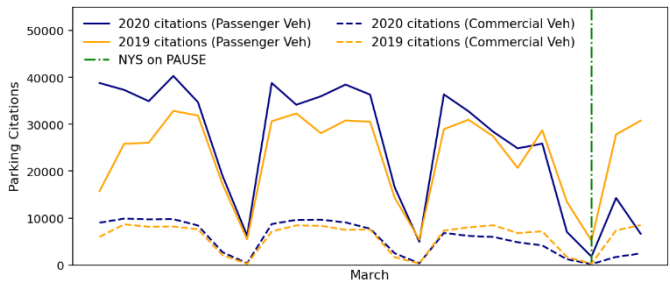

**Figure 6** Parking Citations during COVID-19

## Summary of Findings

One month into the COVID-19 pandemic, the metropolitan area of New York has seen a radical change in travel patterns. Transit ridership and motor vehicle trips are significantly down, and as a result less congestion and fewer reported traffic crashes are observed. Some trip reductions seem to reach the floor at the end of March. While trip reductions have led to faster and safer trips for essential trips, the foregoing declines in transit ridership and vehicular traffic using tolled roadways will also result in massive shortfalls in revenue for transportation agencies. C2SMART researchers are continuing to collect data and monitor these trends as this crisis goes on.

*Note: all data is preliminary and subject to change.*

*For more information please contact c2smart@nyu.edu*